\documentclass[aps,prb,twocolumn,showpacs,preprintnumbers,amsmath,amssymb,superscriptaddress]{revtex4}

\usepackage{graphicx}
\usepackage{dcolumn}
\usepackage{bm}
\usepackage{color}

\newcommand{\ket}[1]{|#1\rangle}

\begin{document}

\title{Transport through single-wall metallic carbon nanotubes in the cotunneling regime}

\author{I. Weymann}
\email{weymann@amu.edu.pl} \affiliation{Department of Physics,
Adam Mickiewicz University, 61-614 Pozna\'n, Poland}

\author{J. Barna\'s}
\affiliation{Department of Physics, Adam Mickiewicz University,
61-614 Pozna\'n, Poland} \affiliation{Institute of Molecular
Physics, Polish Academy of Sciences, 60-179 Pozna\'n, Poland}

\author{S. Krompiewski}
\affiliation{Institute of Molecular Physics, Polish Academy of
Sciences, 60-179 Pozna\'n, Poland}

\date{\today}

\begin{abstract}
Using the real-time diagrammatic technique and taking into account
both the sequential and cotunneling processes, we analyze the
transport properties of single-wall metallic carbon nanotubes
coupled to nonmagnetic and ferromagnetic leads in the full range
of parameters. In particular, considering the two different shell
filling schemes of the nanotubes, we discuss the behavior of the
differential conductance, tunnel magnetoresistance and the shot
noise. We show that in the Coulomb diamonds corresponding to even
occupations, the shot noise becomes super-Poissonian due to
bunching of fast tunneling processes resulting from the dynamical
channel blockade, whereas in the other diamonds the noise is
roughly Poissonian, in agreement with recent experiments. The
tunnel magnetoresistance is very sensitive to the number of
electrons in the nanotube and exhibits a distinctively different
behavior depending on the shell filling sequence of the nanotube.
\end{abstract}

\pacs{72.25.Mk, 73.63.Fg, 85.75.-d, 73.23.Hk }

\maketitle

\section{Introduction}

Transport properties of carbon nanotubes (CNTs) have been a
subject of extensive studies for a few years.
\cite{saito98,anantram06} In particular, very recently there has
been a growing interest in the shot noise measurements of CNTs.
\cite{onac06,wuPRB07,wuPRL07,tsuneta07} The shot noise provides
useful information, not necessarily contained in the current,
about the electronic structure, couplings to external leads, and
indicates the role of correlations and different types of
processes driving the current. \cite{blanterPR00} In addition,
recent experiments on CNTs coupled to ferromagnetic leads have
also shown that nanotubes exhibit a considerable tunnel
magnetoresistance (TMR) effect.
\cite{tsukagoshi99,zhao02,kimPRB02,sahoo05,jensenPRB05,manPRB06,
liuPRB06,nagabhiravaAPL06,krompiewski,cottetPRB06,
cottet06,schonenberger06,kollerNJP07,weymannPRB07} The TMR
provides information about the spin accumulation on the nanotube
and charge states taking part in transport.
\cite{weymannPRB05,weymannPRB07}

If the nanotube is weakly coupled to external leads, the current
flows through the system due to consecutive tunneling processes.
The first-order (sequential) tunneling processes dominate above a
certain threshold voltage and are suppressed below this voltage
due to the single-electron charging energy and/or finite level
separation. On the other hand, current in the blockade regions is
mainly mediated by second-order processes (cotunneling).
\cite{cotunneling} Theoretical considerations of spin-dependent
transport properties of single-wall metallic CNTs in the
perturbative regime have been so far mainly restricted to the
sequential tunneling processes. \cite{kollerNJP07,weymannPRB07}
However, because transport in the Coulomb blockade regime is
dominated by cotunneling processes, the sequential tunneling
approximation may lead to wrong results. Therefore, to get
reliable information on the transport properties in the full range
of bias and gate voltages, which could be compared with that
observed experimentally, one should go beyond the first-order
theory. This is the main objective of the present paper.

The considerations are based on the real-time diagrammatic
technique, which allows us to take into account the sequential
tunneling, cotunneling and cotunneling-assisted sequential
tunneling processes in a fully systematic way. Assuming realistic
parameters of the system, \cite{liangPRL02,sapmazPRB05} we
calculate the current, differential conductance, TMR, and the shot
noise of CNTs. In addition, we also analyze the effect of
different shell filling sequences, which can be realized in the
single-wall CNTs exhibiting four-electron periodicity,
\cite{liangPRL02,sapmazPRB05} on transport characteristics.
Generally, one can distinguish two different shell filling schemes
-- in the first scheme the following sequence of the ground states
is realized: $S=0,1/2,0,1/2$, where $S$ is the spin of the
nanotube, while in the second scheme it is: $S=0,1/2,1,1/2$. We
show that the effect of different shell filling scenarios is the
most visible in the TMR, while the current and shot noise are only
slightly affected. This is due to the fact that the TMR directly
reflects the magnetic properties of the system and it is thus very
sensitive to whether the ground state of the nanotube is a singlet
or a triplet. As concerns the shot noise, we show that the
corresponding Fano factor $F$ in the Coulomb blockade regions is
enhanced above the Schottky value, $F>1$, due to bunching of
inelastic cotunneling processes. However, this enhancement is
significantly lower than that obtained within the first-order
approximation. \cite{weymannPRB07} Contrary, in the transport
regions where the sequential contribution to the current is
dominant, the shot noise is sub-Poissonian with $F$ slightly above
1/2 and is only weakly modified by the cotunneling processes. Our
results for the shot noise are in qualitative agreement with
recent experimental data on carbon nanotubes, \cite{onac06} and
quantum dots. \cite{zhang}

The paper is organized as follows. In section 2 we describe the
model of the nanotube and in section 3 we briefly present the
real-time diagrammatic technique used in calculations. Numerical
results and their discussion is given in section 4, where we first
present the results on CNTs coupled to nonmagnetic leads and then
we discuss the effects of spin-dependent tunneling on transport
properties. In addition, we also present results for two different
shell filling scenarios. Finally, we give the conclusions in
section 5.

\section{Model}

We consider a single wall metallic CNT weakly coupled to the two
electrodes which can be either nonmagnetic or ferromagnetic. In
the latter case the magnetizations of the leads are assumed to be
collinear and form either parallel or antiparallel magnetic
configuration. The Hamiltonian $\hat{H}$ of the system takes the
general form $\hat{H}=\hat{H}_{\rm L} + \hat{H}_{\rm R} +
\hat{H}_{\rm CNT} + \hat{H}_{\rm T}$. The first two terms describe
noninteracting itinerant electrons in the leads, $\hat{H}_r =
\sum_{{\mathbf k}\sigma} \varepsilon_{r{\mathbf k}\sigma}
c^{\dagger}_{r{\mathbf k}\sigma} c_{r{\mathbf k}\sigma}$ for the
left ($r={\rm L}$) and right ($r={\rm R}$) lead, with
$\varepsilon_{r{\mathbf k}\sigma}$ being the energy of an electron
with the wave vector ${\mathbf k}$ and spin $\sigma$ in the lead
$r$. The third term of $\hat{H}$ describes the single-wall CNT and
is given by \cite{oregPRL00}
\begin{eqnarray}\label{Eq:HNT}
   \hat{H}_{\rm CNT} &=& \sum_{\mu j\sigma}
   \varepsilon_{\mu j} n_{\mu j\sigma} + \frac{U}{2}
   \left( N - N_0 \right)^2
   \nonumber\\
   &+& \delta U \sum_{\mu j} n_{\mu j\uparrow} n_{\mu j\downarrow}
   + J \sum_{\mu j, \mu^\prime j^\prime}
   n_{\mu j\uparrow} n_{\mu^\prime j^\prime\downarrow}
   \,,
\end{eqnarray}
where $N = \sum_{\mu j\sigma} n_{\mu j\sigma}$, with $n_{\mu
j\sigma} = d^{\dagger}_{\mu j\sigma}d_{\mu j\sigma}$ being the
occupation operator for spin $\sigma$ and $j$th level in the
subband $\mu$ ($\mu=1,2$). The energy $\varepsilon_{\mu j}$ of the
$j$th discrete level in the subband $\mu$ is given by
$\varepsilon_{\mu j} = j\Delta + (\mu-1)\delta$, where $\Delta$ is
the mean level spacing and $\delta$ is the energy mismatch between
the two subbands of the nanotube. The charging energy of CNT is
described by $U$, and $N_0$ is the charge on the nanotube induced
by gate voltages. The additional on-level Coulomb energy between
two electrons occupying the same orbital level is denoted by
$\delta U$, whereas $J$ describes the exchange energy between the
spin-up and spin-down electrons. The exchange effects described by
$J$ play an important role for small diameter nanotubes
\cite{mayrhoferPRB06} (up to 2 nm or so), as considered in this
paper, whereas for nanotubes of larger diameters these effects
become negligible. Finally, the last term of $\hat{H}$,
$\hat{H}_{\rm T}$, takes into account tunneling processes between
the nanotube and electrodes, $\hat{H}_{\rm T}=\sum_{r=\rm
L,R}\sum_{\mathbf k} \sum_{\mu j\sigma} \left(t_{rj}c^{\dagger}_{r
{\mathbf k}\sigma} d_{\mu j\sigma} + t_{rj}^\star d^\dagger_{\mu
j\sigma} c_{r {\mathbf k}\sigma} \right)$, where $t_{rj}$ denotes
the tunnel matrix elements between the lead $r$ and the $j$th
level (assumed to be spin-independent also for ferromagnetic
leads). Coupling of the $j$th level to external leads is described
by $\Gamma_{rj}^{\sigma}= 2\pi |t_{rj}|^2 \rho_r^\sigma$, with
$\rho_r^\sigma$ being the density of states in the lead $r$ for
spin $\sigma$. The role of ferromagnetic leads is taken into
account just {\it via} the spin-dependent density of states
$\rho_r^\sigma$. The coupling parameters can also be expressed as
$\Gamma_{rj}^{+(-)}=\Gamma_{rj}(1\pm p_{r})$, for the
spin-majority (spin-minority) electron bands, with $\Gamma_{rj}=
(\Gamma_{rj}^{+} +\Gamma_{rj}^{-})/2$ and $p_r$ being the spin
polarization of the lead $r$, $p_{r}=(\rho_{r}^{+}- \rho_{r}^{-})/
(\rho_{r}^{+}+ \rho_{r}^{-})$. For nonmagnetic leads,
$\Gamma_{rj}^{+}=\Gamma_{rj}^{-}$. In the following we assume
$\Gamma_{rj}=\Gamma/2$ for all values of $j$ and $r$.

\section{Method}

In order to calculate the transport through a single-wall metallic
carbon nanotube in the sequential and cotunneling regimes, we
employ the real-time diagrammatic technique.
\cite{weymannPRB07,diagrams,thielmannPRL05} It consists in a
systematic expansion of the nanotube (reduced) density matrix and
the operators of interest with respect to the coupling strength
$\Gamma$. The time evolution of the reduced density matrix can be
visualized as a sequence of irreducible self-energy blocks,
$\mathbf{W}$, on the Keldysh contour. The matrix elements $W_{\chi
\chi^\prime}$ of $\mathbf{W}$ describe transitions between the
many-body states $\ket{\chi}$ and $\ket{\chi^\prime}$ of the CNT.
\cite{weymannPRB07} Then, the Dyson equation constitutes the full
propagation of the reduced density matrix, which leads to a
kinetic equation for the nanotube occupation probabilities,
$(\mathbf{\tilde{W}}\mathbf{p}^{\rm st})_{\chi} =
\Gamma\delta_{\chi\chi_0}$, where $\mathbf{p}^{\rm st}$ is the
vector containing probabilities and the matrix
$\mathbf{\tilde{W}}$ is given by $\mathbf{W}$ with one arbitrary
row $\chi_0$ replaced by $(\Gamma,\dots,\Gamma)$ due to the
normalization condition $\sum_{\chi}p_{\chi}^{\rm st}=1$. The
current flowing through the system can be found from
$I=(e/2\hbar){\rm Tr}\{\mathbf{W}^{\rm I}\mathbf{p}^{\rm st}\}$,
whereas the zero-frequency current noise, $S=2\int_{-\infty}^0 dt
( \langle \hat{I}(t)\hat{I}(0)+\hat{I}(0)\hat{I}(t)\rangle-2
\langle \hat{I}\rangle^2)$, is given by \cite{thielmannPRL05} $S =
(e^2/\hbar){\rm Tr}\left\{\left[ \mathbf{W}^{\rm II}+
\mathbf{W}^{\rm I} \left( \mathbf{P}\mathbf{W}^{\rm I} +
\mathbf{p}^{\rm st} \otimes \mathbf{e}^{\rm T}\partial
\mathbf{W}^{\rm I}\right)\right] \mathbf{p}^{\rm st}\right\}$. The
matrix $\mathbf{W}^{\rm I (II)}$ is the self-energy matrix with
one {\it internal} vertex (two internal vertices), resulting from
the expansion of the tunneling Hamiltonian, replaced by the
current operator, while the matrices $\partial\mathbf{W}$ and
$\partial\mathbf{W}^{\rm I}$ are partial derivatives of
$\mathbf{W}$ and $\mathbf{W}^{\rm I}$ with respect to the
convergence factor of the Laplace transform. \cite{thielmannPRL05}
The object $\mathbf{P}$ is calculated from:
$\mathbf{\tilde{W}}\mathbf{P} = \mathbf{\tilde{1}}(\mathbf{p}^{\rm
st}\otimes \mathbf{e}^{\rm T} - \mathbf{1} -
\partial\mathbf{W}\mathbf{p}^{\rm st}\otimes \mathbf{e}^{\rm T})$,
with $\mathbf{\tilde{1}}$ being the unit vector with row $\chi_0$
set to zero, and $\mathbf{e}^{\rm T} = (1,\dots,1)$.
\cite{thielmannPRL05} In order to calculate the transport
properties order by order in tunneling processes, we expand the
self-energy matrices, $\mathbf{W}^{\rm (I,II)} = \mathbf{W}^{\rm
(I,II)(1)} +\mathbf{W}^{\rm (I,II)(2)}+\dots$, the dot
occupations, $\mathbf{p}^{\rm st} = \mathbf{p}^{\rm st (0)} +
\mathbf{p}^{\rm st (1)}+\dots$, and, $\mathbf{P} =
\mathbf{P}^{(-1)} + \mathbf{P}^{(0)}+\dots$, respectively. The
self-energies can be calculated using the corresponding
diagrammatic rules. \cite{thielmannPRL05,weymannPRB05,weymann}

\section{Numerical results and discussion}

The shell filling structure of single-wall carbon nanotubes
exhibits the four-electron periodicity, which is a consequence of
the two spin-degenerate subbands of the nanotube. Furthermore,
depending on the intrinsic parameters of the nanotube, its shell
filling can generally have two different scenarios.
\cite{liangPRL02,sapmazPRB05} The first one is associated with the
following sequence of the ground states: $S=0,1/2,0,1/2$, which is
realized when $\delta U+J<\delta$. On the other hand, in the
second scenario the spin state of the nanotube changes as
$S=0,1/2,1,1/2$, which happens for $\delta U+J>\delta$. The
difference between these two cases appears in the transport regime
where the nanotube is doubly occupied. In the case of first
sequence, i.e. for $\delta U+J<\delta$, one orbital level of one
of the subbands is fully occupied and the nearest orbital level of
the other subband is empty - this is a singlet state. Whereas for
$\delta U+J>\delta$, there is a single electron on each level of
the two subbands which, due to the exchange interaction $J$, lead
to the formation of a triplet state. In the following, we analyze
the behavior of the tunnel magnetoresistance and the shot noise in
the linear and nonlinear response regimes depending on two
different shell filling schemes.

\subsection{Shell filling sequence: $S=0,\frac{1}{2},0,\frac{1}{2}$}

\begin{figure}[t]
  \includegraphics[width=0.8\columnwidth]{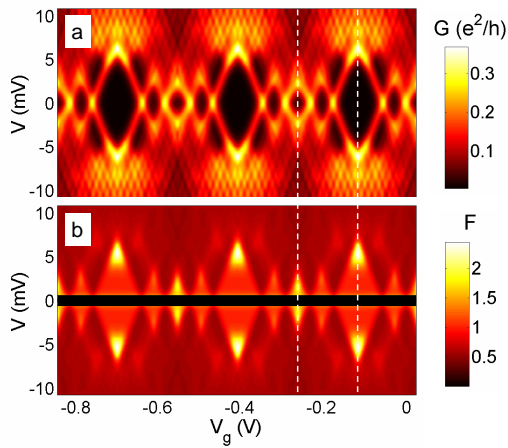}
  \caption{\label{Fig:1} (Color online)
  Differential conductance (a) and Fano factor (b)
  as a function of bias and gate voltages.
  The parameters are: $\Delta = 8.4$ meV, $U/\Delta = 0.26$,
  $J/\Delta = 0.12$, $\delta U/\Delta=0.04$,
  $\delta/\Delta = 0.27$, $k_{\rm B}T/\Delta = 0.035$,
  $p_{\rm L} = p_{\rm R} = 0$, $x=0.14$, and $\Gamma = 0.2$ meV.}
\end{figure}

When $\delta U+J<\delta$, the shell filling of the nanotube is
realized in this way that the next orbital levels are being
occupied only after the lower lying levels are full, which leads
to the sequence of doublet and singlet ground states,
$S=0,1/2,0,1/2$. To model the nanotube in this transport regime,
we have taken the experimental parameters derived by W. Liang {\it
et al.} \cite{liangPRL02} Furthermore, to clearly elucidate the
effect of cotunneling on transport properties, we first present
and discuss the behavior of the differential conductance and the
Fano factor in the case of CNT coupled to nonmagnetic leads. In
this case, for comparison we also show the results obtained when
taking into account only the first-order tunneling.

In Fig.~\ref{Fig:1} we show the density plots of the differential
conductance (a) and the Fano factor (b) as a function of the bias
and gate voltages. Since the noise in the small bias regime
($|eV|\le k_{\rm B}T$), is dominated by thermal Nquist-Johnson
noise, we excluded this part from considerations and marked it by
the black thick line around $V=0$ (the corresponding Fano factor
diverges as $V \to 0$). The black diamonds in Fig.~\ref{Fig:1}(a)
correspond to blockade regions, where sequential transport is
exponentially suppressed and current is dominated by the second
order processes. Furthermore, the four-electron periodicity is
clearly visible in the linear conductance, while the additional
lines in the differential conductance for voltages larger then the
threshold for sequential tunneling are due to excited states
participating in transport.

\begin{figure}[t]
 \includegraphics[width=0.49\columnwidth]{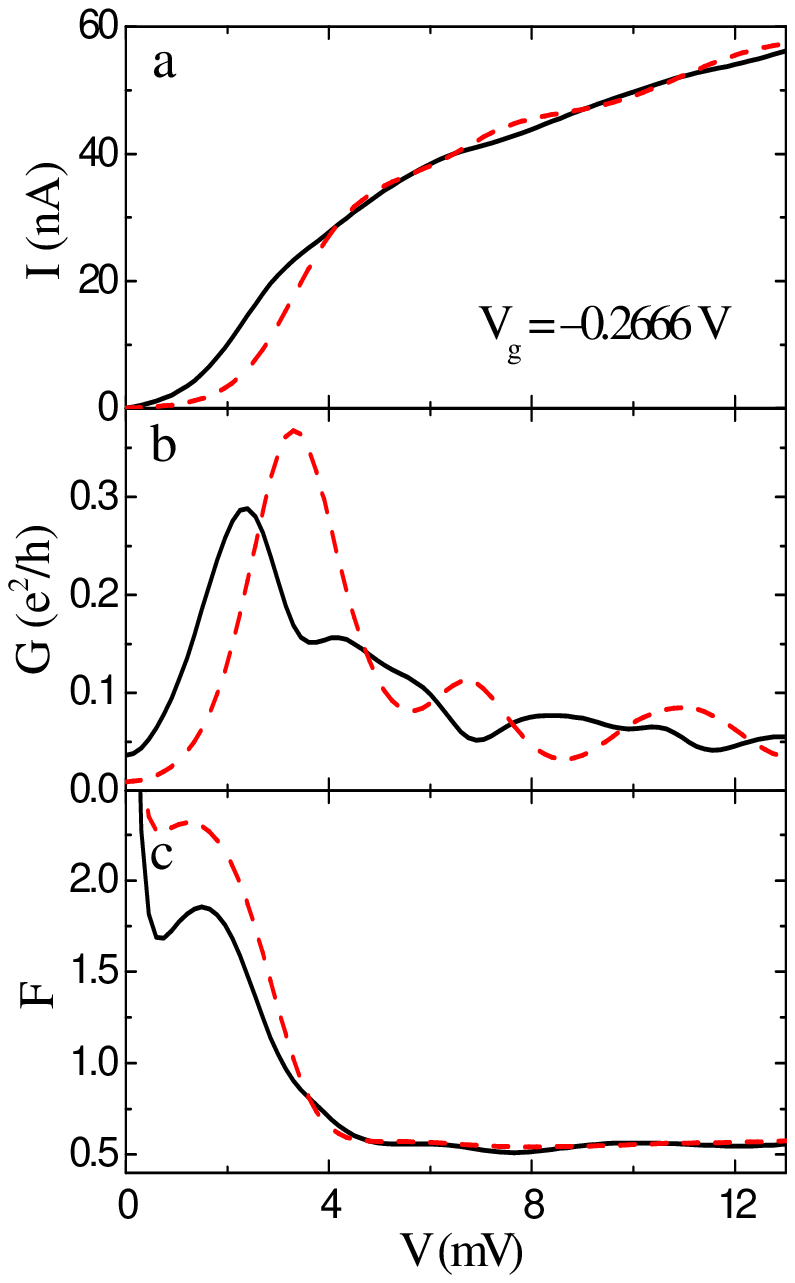}
 \includegraphics[width=0.49\columnwidth]{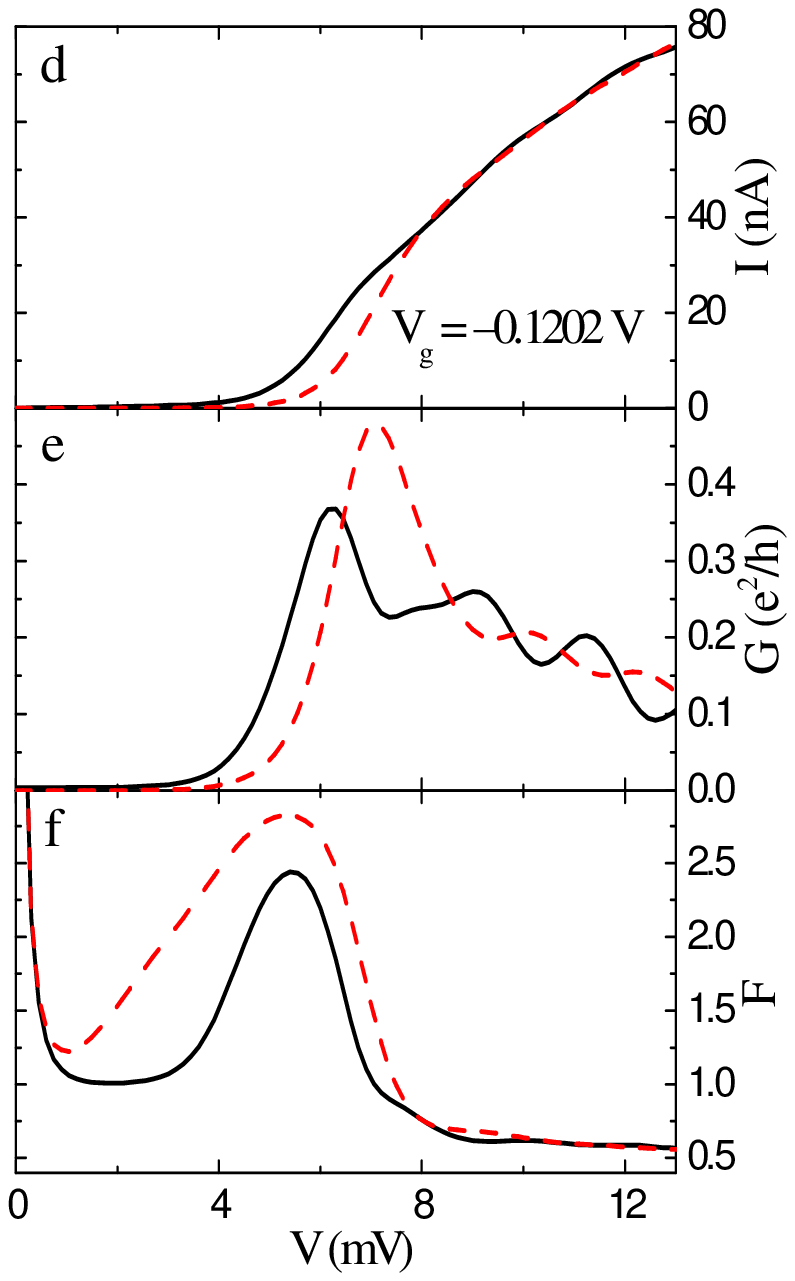}
  \caption{\label{Fig:2}
  (Color online) Bias dependence of the current,
  differential conductance
  and the Fano factor for $V_g = -0.2666$ V [(a)-(c)]
  and $V_g = -0.1202$ V [(d)-(f)], see the dashed lines in Fig.~\ref{Fig:1}.
  The dashed lines show the first-order calculation.
  The parameters are the same as in Fig.~\ref{Fig:1}.}
\end{figure}

To see clearly the contribution from cotunneling processes we show
in Fig.~\ref{Fig:2} the current (a), conductance (b), and Fano
factor (c) for two different values of the gate voltage,
corresponding to vertical cross-sections in Fig.~\ref{Fig:1}
through small and large Coulomb diamonds, see the dashed lines in
Fig.~\ref{Fig:1}. Additionally, in Fig.~\ref{Fig:2} we also show
the corresponding quantities calculated in the first order (dashed
lines). It is evident that the cotunneling processes modify the
current mainly in the blockade regime and also close to resonance
(blockade threshold), leading to renormalization of the nanotube
levels. This makes the conductance peak shifted towards smaller
voltages. As regards the shot noise, cotunneling processes have a
significant impact on the Fano factor, particularly in the
blockade regions corresponding to even occupations, where the shot
noise is super-Poissonian. The corresponding Fano factor is then
larger than unity, but significantly reduced in comparison to that
calculated in the first order approximation.

Consider first the limit of sequential (first order) tunneling
processes. The sequential tunneling starts when the bias
approaches the threshold voltage. Owing to the Fermi distribution,
sequential processes start before the zero-temperature threshold
is reached. When the highest populated level is doubly occupied,
tunneling from the source to the drain electrode may leave the
nanotube either in the initial or in an excited state. When the
CNT is left in the initial state, the next tunneling process goes
with the same probability as the first one. When, however, the CNT
is left in an excited state, the thermal (dynamical) blockade is
lifted and the subsequent tunneling processes are faster. This
leads to bunching of fast tunneling processes, which leads to the
super-Poissonian noise. \cite{bulka00,belzigPRB04,
weymannJPCM07,cottet04,sukhorukovPRB01}

The cotunneling processes, both single-barrier and two-barrier
ones, are not blocked. Generally, there are elastic and inelastic
ones, which occur with different probabilities. However, the
asymmetry between the fast and slow ones is reduced in comparison
with the first order, which leads to a smaller Fano factor when
the second order processes are included. Apart from this, the
inelastic cotunneling processes also lift the dynamical blockade
of the sequential processes.

\begin{figure}[t]
  \includegraphics[width=0.8\columnwidth]{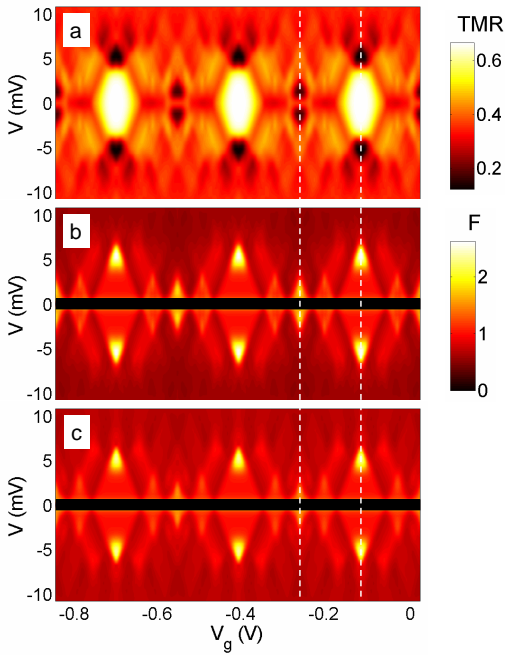}
  \caption{\label{Fig:3} (Color online)
  Tunnel magnetoresistance (a) and Fano factor
  in the parallel (b) and antiparallel (c) configuration
  as a function of bias and gate voltages.
  The parameters are as in Fig.~\ref{Fig:1} with
  $p_{\rm L} = p_{\rm R} = 0.5$.}
\end{figure}

It is easy to note, that the above scenario does not hold when the
highest level is singly occupied. In such a case first order
tunneling processes are of comparable velocity and therefore the
Fano factor is only slightly above unity. This applies also to
cotunneling processes. Thus, depending on the ground state of the
nanotube which can be changed by the gate voltage, we either find
a considerable enhancement of the Fano factor (super-Poissonian
shot noise) or the Fano factor is approximately equal unity
(Poissonian shot noise). This observation is in agreement with
recent experiments carried out by E. Onac {\it et al.}
\cite{onac06} Finally, it is noteworthy that outside the blockade
regions the shot noise becomes sub-Poissonian and the corrections
due to cotunneling processes are much smaller than in the blockade
regime, although still noticeable.

\begin{figure}[t]
  \includegraphics[width=0.49\columnwidth]{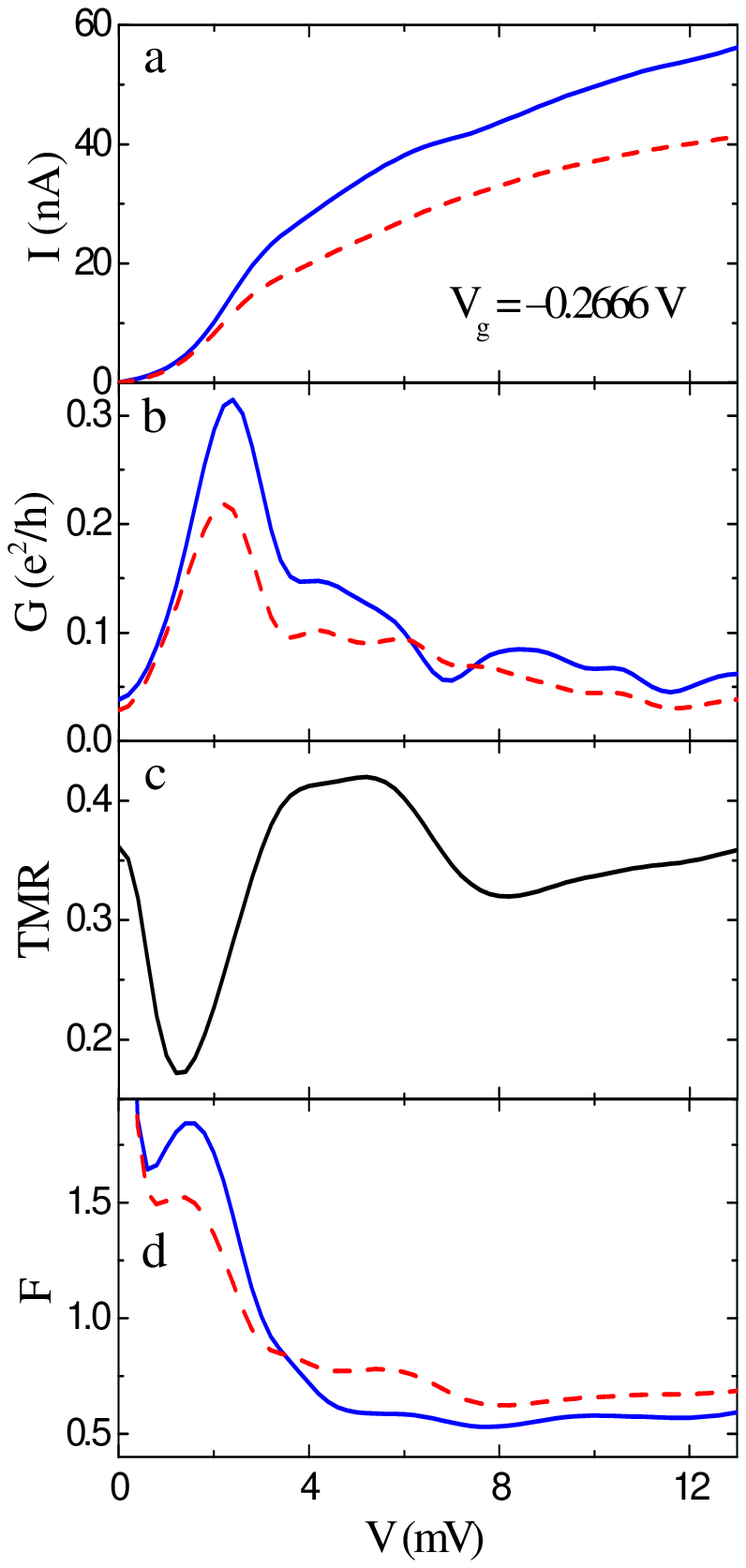}
  \includegraphics[width=0.49\columnwidth]{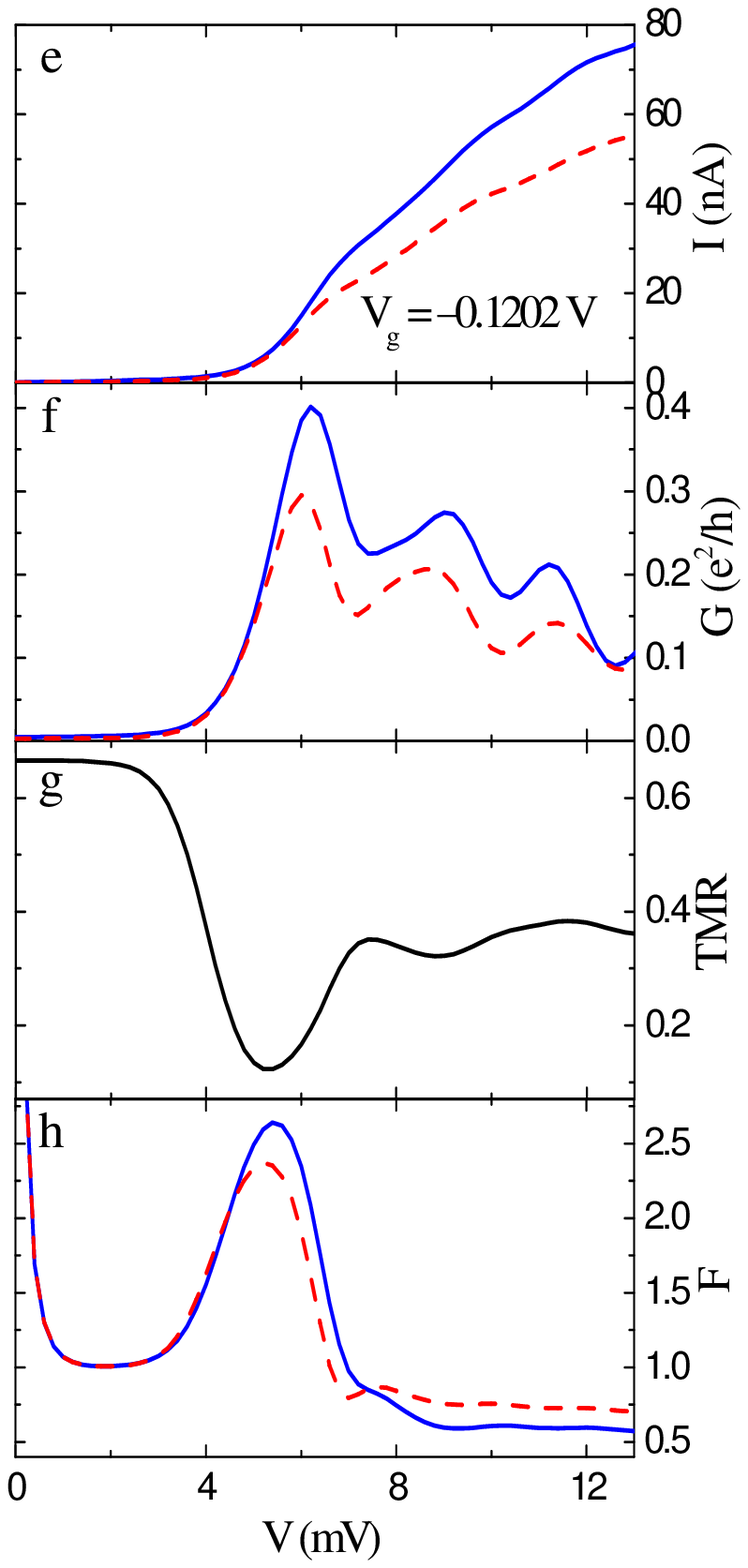}
  \caption{\label{Fig:4}
  (Color online) Bias dependence of the current,
  differential conductance, TMR
  and the Fano factor in the parallel (solid line)
  and antiparallel (dashed line) configuration
  for $V_g = -0.2666$ V [(a)-(d)]
  and $V_g = -0.1202$ V [(e)-(h)],
  see the dashed lines in Fig.~\ref{Fig:3}.
  The parameters are the same as in Fig.~\ref{Fig:3}.}
\end{figure}

Consider now the numerical results on CNTs contacted to
ferromagnetic leads, see Fig.~\ref{Fig:3}. The tunnel
magnetoresistance, Fig.~\ref{Fig:3}(a), exhibits a distinctively
different behavior depending on the number of electrons on the
nanotube. In the linear response regime, between the two
consecutive four-electron sequences, the TMR is given by the
Julliere's formula, ${\rm TMR} = 2p_{\rm L}p_{\rm R}/(1- p_{\rm
L}p_{\rm R})$, \cite{julliere75} whereas in the other transport
regimes it is suppressed below the Julliere's value. This is a
completely new feature as compared to the sequential tunneling
results where the linear response TMR is constant and given by
${\rm TMR _{seq}} = p_{\rm L}p_{\rm R}/(1- p_{\rm L}p_{\rm R})$,
irrespective of the gate voltage. \cite{kollerNJP07,weymannPRB07}

In Fig.~\ref{Fig:3} (b) and (c) we show the density plots of the
Fano factors in both parallel and antiparallel magnetic
configurations. The general features of the Fano factors are
similar to those in the limit of nonmagnetic leads, except for the
fact that now there is an additional contribution to the shot
noise coming from the difference between the two spin channels for
transport. Because the resultant coupling of the nanotube to
ferromagnetic leads in the parallel configuration is larger than
in the antiparallel one, the fluctuations of the current are
enhanced in the parallel configuration as compared to the
antiparallel one. This leads to the corresponding difference in
the Fano factors, see Fig.~\ref{Fig:3} (b) and (c), which is
especially visible in the Coulomb blockade regime. We also note
that although the magnitudes of the Fano factors in both magnetic
configurations are different, their general behavior is
qualitatively similar.

The bias dependence of the current, conductance, TMR and the Fano
factor is shown in Fig.~\ref{Fig:4} for two different values of
the gate voltage and for both the parallel and antiparallel
magnetic configurations. All the quantities are calculated taking
into account both the sequential and cotunneling processes, and we
do not distinguish between the first and second order
contributions. As in the case of nonmagnetic leads, the second
order processes significantly modify the TMR and Fano factor in
the blockade regions, and only slightly outside these regions. An
interesting feature is a drop of the TMR at the onset for
sequential tunneling, see Fig.~\ref{Fig:4}(c) and (g), which is
present when the ground state of the nanotube is a singlet. This
is due to the interplay of the inelastic cotunneling processes
which flip the spin of electrons occupying the nanotube and the
sequential tunneling which is generally incoherent. On the other
hand, when the ground state of the nanotube is a doublet, the TMR
does not drop with approaching the threshold, see
Fig.~\ref{Fig:3}(a), which is due to a nonequilibrium spin
accumulation induced in the nanotube. The spin accumulation tends
to increase the TMR and, thus, reduces the role of the
above-mentioned processes responsible for TMR drop in the case of
ground state $S=0$.

\begin{figure}[t]
  \includegraphics[width=0.8\columnwidth]{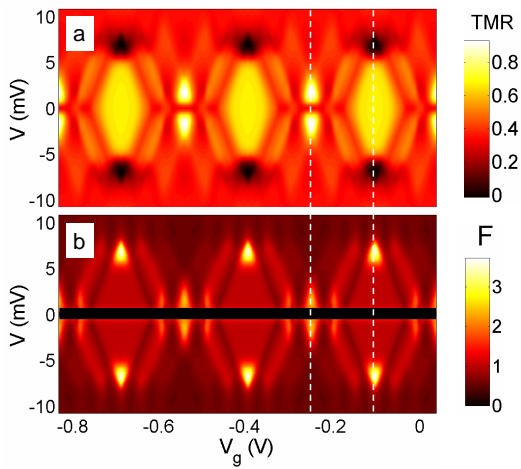}
  \caption{\label{Fig:5} (Color online)
  Tunnel magnetoresistance (a) and Fano factor
  in the parallel (b) configuration
  as a function of bias and gate voltages.
  The parameters are the same as in Fig.~\ref{Fig:3}
  with $\delta/\Delta = 0.1$.}
\end{figure}

\subsection{Shell filling sequence: $S=0,\frac{1}{2},1,\frac{1}{2}$ }

The intrinsic parameters of the nanotube depend mainly on their
size and coupling to external leads. \cite{liangPRL02,sapmazPRB05}
In particular, the quality of the coupling between the leads and
nanotube contributes to the energy mismatch $\delta$ between the
two CNT's subbands. In the case when $\delta U+J>\delta$, the
sequence of the ground states is changed as compared to the case
of $\delta U+J<\delta$. Now, in the charge state with two
electrons in the nanotube, each electron occupies one orbital
level of the two subbands. This, together with a finite $J$, leads
to the formation of a triplet state in the nanotube. Thus, the
shell filling sequence becomes $S=0,1/2,1,1/2$.

\begin{figure}[t]
  \includegraphics[width=0.49\columnwidth]{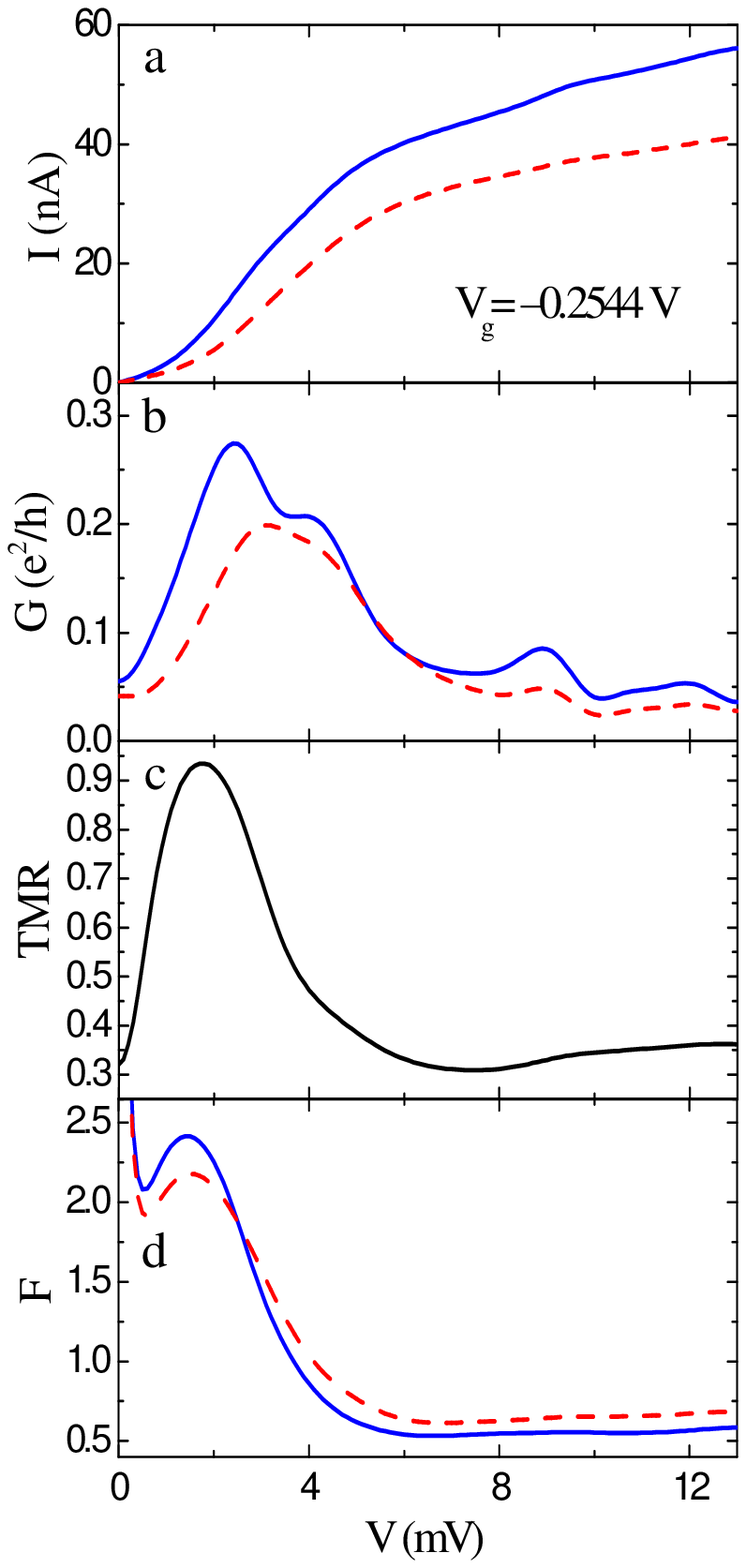}
  \includegraphics[width=0.49\columnwidth]{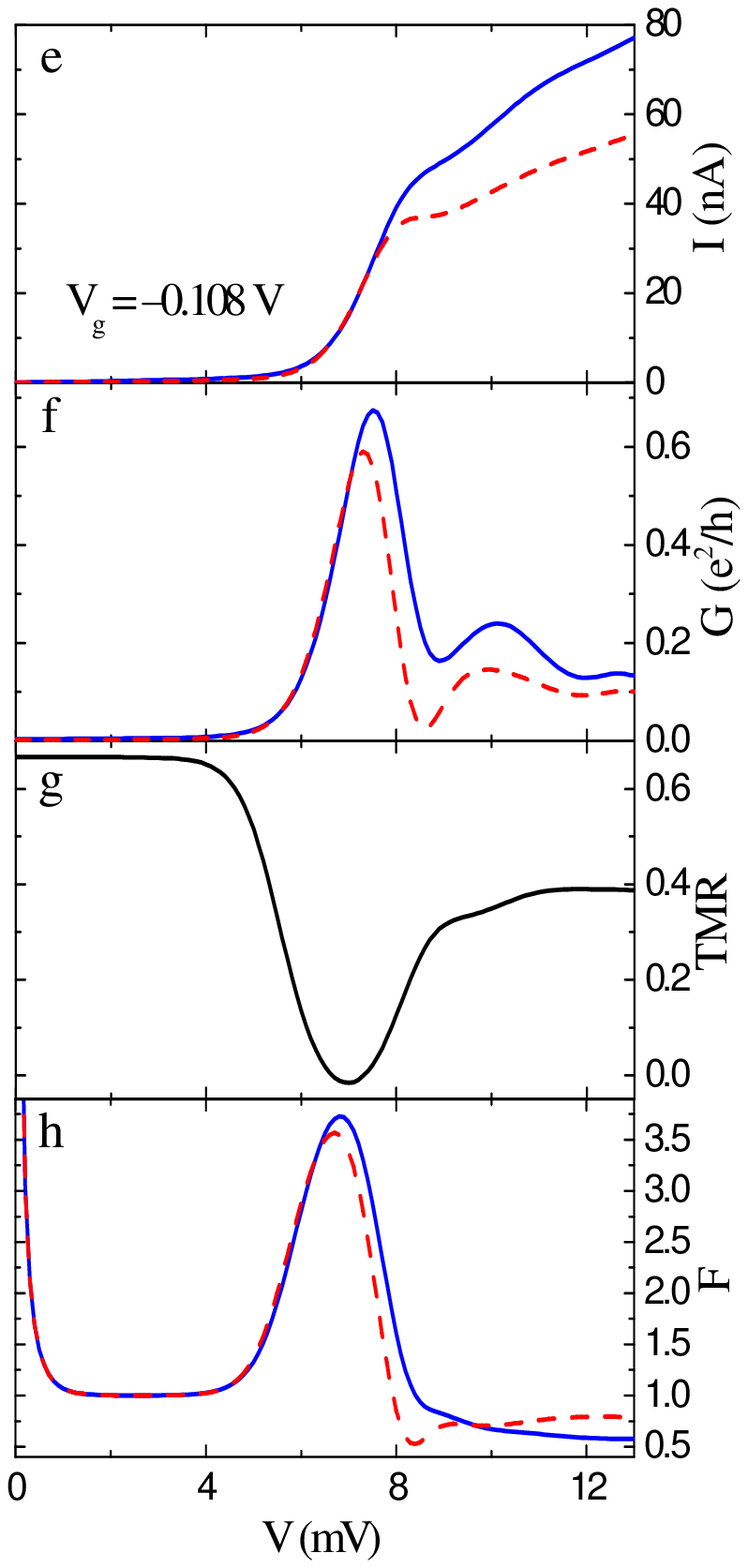}
  \caption{\label{Fig:6}
  (Color online) Bias dependence of the current,
  differential conductance, TMR
  and the Fano factor in the parallel (solid line)
  and antiparallel (dashed line) configuration
  for $V_g = -0.2544$ V [(a)-(d)]
  and $V_g = -0.108$ V [(e)-(h)],
  corresponding to the dashed lines in Fig.~\ref{Fig:5}.
  The parameters are the same as in Fig.~\ref{Fig:5}.}
\end{figure}

In Fig.~\ref{Fig:5} we show the TMR and the Fano factor in the
parallel configuration as a function of the bias and gate
voltages. The Fano factor in the antiparallel configuration is
generally smaller than that in the parallel one, but its behavior
is qualitatively similar, therefor we only show $F$ in the
parallel configuration -- it is displayed in Fig.~\ref{Fig:5}(b).
Furthermore, it can be seen that the general behavior of the Fano
factor is similar to that in the case of $\delta U+J<\delta$, see
Figs.~\ref{Fig:1}(b) and \ref{Fig:3}(b). Now, the Fano factor
becomes super-Poissonian when the ground state of the nanotube is
either a singlet or a triplet. This fact is rather intuitive as
the dependence of the shot noise on the bias and gate voltages is
rather conditioned by the charge states taking part in transport
than by the magnetic state of the system. The information about
the spin state of the nanotube is mainly contained in the TMR,
which for $\delta U+J>\delta$ becomes considerably modified,
especially in the transport regime where the ground state is a
triplet, see Fig.~\ref{Fig:5}(a).

The bias voltage dependence of the current, differential
conductance, Fano factor in the parallel and antiparallel
configurations as well as of the TMR is shown in Fig.~\ref{Fig:6}.
One can now clearly see that general features of transport
characteristics calculated for $V_g = -0.2544$ V, i.e. when the
ground state of the nanotube is a singlet, are similar to those
shown in Fig.~\ref{Fig:4}(a)-(d). The only difference is the
enhancement of the Fano factor at resonance, see
Fig.~\ref{Fig:6}(f), and the drop of the TMR at the onset of
sequential tunneling where now TMR becomes slightly negative, see
Fig.~\ref{Fig:6}(g). On the other hand, when the ground state of
the nanotube is a triplet ($V_g = - 0.108$ V), the transport
characteristics become even more modified, see
Figs.~\ref{Fig:4}(e)-(g) and \ref{Fig:6}(e)-(g). The main
difference is visible in the behavior of the TMR which for $\delta
U+J>\delta$ is increased above the Julliere's value. This is
associated with the nonequilibrium spin accumulation in triplet
states, which tends to enhance the difference between the two
magnetic configurations of the system. Furthermore, it is also due
to a finite exchange interaction $J$, because of which only
particular components of triplet take part in transport. In this
way, the exchange interaction plays a role similar to an external
magnetic field. As already shown in the case of quantum dots, the
TMR may be enhanced above the Julliere's value when only
particular spin states of the dot participate in transport.
\cite{weymannJPCM07}

\section{Concluding remarks}

In conclusion, we have calculated transport characteristics,
including shot noise and tunnel magnetoresistance, of CNTs weakly
coupled to magnetic and nonmagnetic leads. We have also analyzed
the effects of two different shell filling scenarios, which can be
realized in single-wall metallic CNTs, on transport
characteristics. In particular, we have shown that the second
order contributions are crucial in the blockade regions. Depending
on the ground state of the nanotube, the shot noise is either
supper-Poissonian due to the bunching of fast tunneling processes,
or approximately given by the Poisson value, in agreement with
recent experiments. \cite{onac06} On the other hand, in the
regions dominated by sequential transport processes the shot noise
is sub-Poissonian with the corresponding Fano factor roughly equal
to 1/2. Furthermore, we have also shown that the linear TMR
exhibits a strong dependence on the gate voltage, being equal to
the Julliere's TMR in transport regime between the two consecutive
four-electron fillings and suppressed below that value in the
other transport regimes. In addition, when the ground state of the
nanotube is a triplet, the TMR in the nonlinear response regime at
the onset for sequential tunneling becomes enhanced above the
Julliere's value, which is due to nonequilibrium spin accumulation
in triplet states and finite exchange interaction $J$.

\begin{acknowledgments}

This work was supported by the EU grant CARDEQ under contract
IST-021285-2 and, as part of the European Science Foundation
EUROCORES Programme SPINTRA (contract No. ERAS-CT-2003-980409), by
funds from the Ministry of Science and Higher Education as a
research project in years 2006-2009. I.W. also acknowledges
support from the Foundation for Polish Science.

\end{acknowledgments}

\end{document}